# Eye of the Mind: Image Processing for Social Coding

Maleknaz Nayebi
Ecole Polytechnique of Montreal
Montreal, Canada
mnayebi@polymtl.ca

## ABSTRACT

Developers are increasingly sharing images in social coding environments alongside the growth in visual interactions within social networks. The analysis of the ratio between the textual and visual content of Mozilla's change requests and in Q/As of StackOverflow programming revealed a steady increase in sharing images over the past five years. Developers' shared images are meaningful and are providing complementary information compared to their associated text. Often, the shared images are essential in understanding the change requests, questions, or the responses submitted. Relying on these observations, we delve into the potential of automatic completion of textual software artifacts with visual content.

## 1 INTRODUCTION

The use of images on social media is not restricted to the general public. The visual content made the day to day communications easier (sending the photo of your food instead of writing its long name) and made the image-based social networks, such as Snapchat and Instagram, the new trend. The impact of social networks on development environments is known [11]. Sharing images can also help developers to communicate more effectively as a picture worth a thousand words. The use of diagrams and brainstorming on office boards is a known part of the work in software teams. However, the existing social coding platforms and the software engineering automated tools are unable to retrieve, process, and extract information from developers' shared images. Motivated by the advances in deep neural networks and image processing [3, 12], and by increasing attention to visual content in social media and chat clients, we envision the potential of using images for communication between developers, e.g., reporting bugs or asking and answering development questions. We support this vision through three claims and provide empirical evidence.

**1- Increasing trend of sharing images:** There is a an increasing trend in visual content shared by developers directly related to software development processes and tasks (Section 2).

**2- Images complement text:** Images provide additional information in social coding environments comparing to the text (Section3).

**3- Images facilitate developers communication:** Images help developers to react faster and understand better (Section4).

Software development is a communication-intensive process that requires understanding, synchronization, and discussion of information. So far, the evolution of tools and techniques for assisting software developers was dominated by understanding and generating only the textual content. The growing number of images shared by developers provides a unique opportunity to train machines on reusing and even synthesizing new images to make the software artifacts visual. **We envision that automated tools will not only extend to *process posted images* along with natural and programming language processing, but also to *generate visual artifacts* to assist developers.** In the past, we could successfully summarize software artifacts (bug reports, source code, mailing lists, developer discussions) into shorter text [8]. We went further and automatically generate textual commit messages, release notes, pull requests or reply to StackOverflow questions, and built chatbots by relying on analogy and natural language processing techniques [1, 4, 9]. In the future, the developers' knowledge sharing and communication assistance tools (such as the ones named above) will complement the current text-based artifacts by synthesizing and generating images. These images would be either retrieved and reused from shared images or will be synthesized and newly generated by automated techniques.

This future is accessible and mainly encouraged by the success in teaching machines sophisticated features such as humans' facial and body attributes, or attributes of transportation means and driveways for self-driving cars. Hence, it is quite probable that automated tools for helping software developers achieve high precision on generated images. Software interfaces, inputs, and outputs are less geometrically complex and with less variety in comparison to solved problems in the domain of image processing in self-driving cars. Hence, machines can learn and synthesize software and development related images. We discuss this future in Section 5.

## 2 INCREASING TREND OF SHARING IMAGES

Instagram and Pinterest are prominent examples of visual social networks. While limiting the size of the textual content, their main intent is to share images. Considering the social nature of software development and its impact on development practices [11], it can be expected that developers would lean toward repeating the behavior met in other social medias. StackOverflow and Bugzilla allow users to attach additional files along with their posted content. Taking these two channels as a widely acknowledged medium for software maintenance and knowledge management activities, we examined the trend of sharing visual content in these two channels. We used Stackexchange data[1] between 2013 to 2018 to mine posts with images. We gathered change requests and replies in Mozilla's Core, Firefox, Firefox for Android, and Firefox for iOS projects[2] using the Bugzilla's REST API. We use this dataset through out the paper.

For the period of 2013 to 2018, StackOverflow included 1,153,318 threads associated with one of the tags Python, Java, or JavaScript. These treads included 2,934,234 posts overall. Figure 2 shows the number of posts with/without images and the ratio of the number of posts, including an image of the total number of posts per year. This analysis has been shown for the data overall and per programming language. The results show a linearly increasing trend in the number of posts, including an image on StackOverflow. Also, our analysis showed (linearly) an increasing trend in the proportion of the number of questions, including an image in StackOverflow. The trend of sharing images in Bugzilla follows the same pattern of steady increase. Figure 3 shows the number of change requests and replies (we refer to them as "posts") for four Mozilla projects between 2013 to 2018. Looking into the number of posts with images relative to the number of all the posts on Bugzilla (26,808 posts) showed an increasing trend over the past five years. Only considering the main change requests (not the follow-up replies and posts), the trend is increasing in a way that in 2018, the number of change requests with images doubled (200% increase) while the total number of change requests only increased by 20%. Mozilla's iOS application was launched in 2015 and there were no change requests before that time.

---

[1] https://archive.org/details/stackexchange
[2] https://bugzilla.mozilla.org/describecomponents.cgi

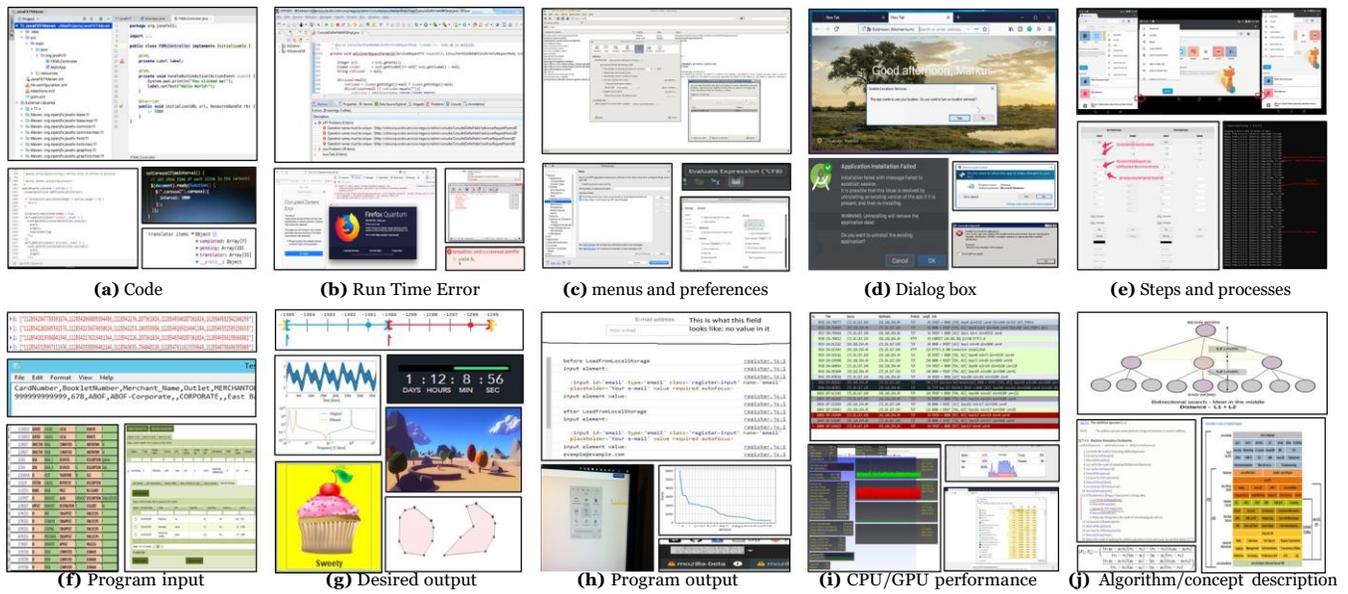

**Figure 1: Sample images posted by developers in Bugzilla or StackOverflow (further details on Section 3 and Table 1).**

## 3 IMAGES COMPLEMENT TEXT

Observing that there is an increasing tendency on sharing images by developers, we were interested to explore if these images are informative and essential to understand the posted text (either the Q/A or the change request). Hence, for 10% of randomly selected (17,374) images from StackOverflow and 50% of randomly selected (5,187) images from Bugzilla, we performed crowdsourced evaluation with 168 software developers. We provided the text (textual body of the StackOverflow questions/replies or the bug reports) and asked developers **(Q1)** if the image provides additional information to the text (Yes/No question) and **(Q2)**, how likely one would understand the text without having the image (Likert scale).

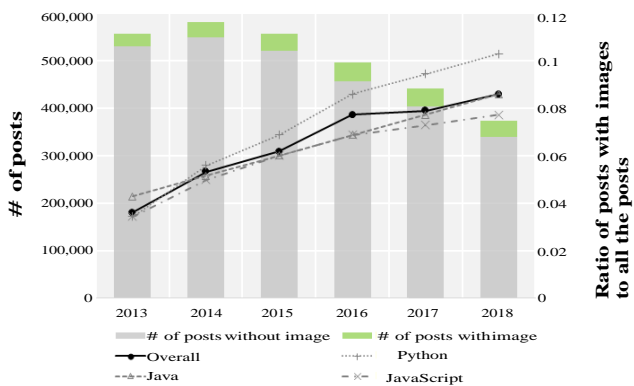

**Figure 2: Trend of sharing images in StackOverflow between 2013 to 2018 for three popular programming languages.**

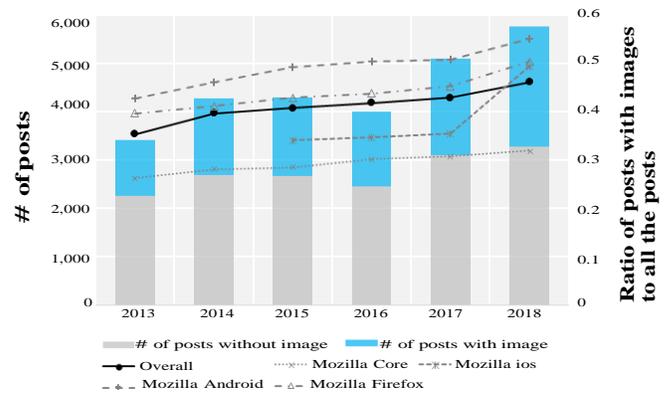

**Figure 3: Trend of sharing images in four popular Mozilla projects between 2013 to 2018.**

We answered (Q1) and (Q2) by showing each image to *three* software developers. We hired these developers by posting a task on Amazon Mechanical Turk[3]. We opened the task only to the developers having an access code posted on BountySource[4]. Overall, 168 developers participated in our crowdsourced tasks for evaluating 22,561 images. These developers had a minimum of two years professional experience and have been active in open source communities. Table 1 summarizes the results of the analysis based on the majority of developers' votes per image. The majority of the images (87.8%) have been identified as informative, and developers considered 86.9% of the texts unlikely or very unlikely to comprehend without the image.

Further, we evaluated the nature of the shared images **(Q3)**. We hired three developers with minimum of seven years coding experience, to identify categories for the content of images. Each developer answered "Q3: What does the image communicate?" for 2,000 randomly selected images (evenly distributed across Bugzilla/StackOverflow, years, and programming language/product). Developers could openly define and name the categories of what the image is communicating and could categorize each image in multiple classes. Then we asked the developers to peer review the categories of one other developer and revise their own categorization if needed (average Cohen's Kappa agreement = 0.83). Finally, we reviewed all categories and aggregated them to ten categories with highest frequency of occurrence. The results are shown in Figure 1.

Having these categories, we started a second crowdsourcing task by asking developers to categorize 22,561 images (10% of StackOverflow posts and 50% of Bugzilla change requests) in at least one of the ten categories defined or categorize them as "other". Table 1 shows the categories and the frequency of images within each category. Overall, the majority (62.8%) of the images contained *code*. Many images categorized as *run time error* or the *menu* often included code as well. The proportion of these images was different between StackOverflow and Mozilla. We found that the majority of the images along with change requests (Mozilla's change requests) are related to *run time errors* while *Code* has the highest percentage on StackOverflow.

## 4 IMAGES FACILITATE COMMUNICATION

We asked the 168 developers why they use images and how they perceive the value of sharing images within social coding environments **(Q4)**-Table 1. The majority of developers (89.9%) use images to help to comprehend the text. 82.7% also used it to be faster in describing and posting. 70.8% of the developers considered images useful for being more precise about the details. 10.1% selected other reasons and described it as *being more fun*, and *being catchy*.

As a more specific type of analysis, we also looked into the time takes to close a change requests on Bugzilla or questions on StackOverflow (for the data described in Section 2). $Time_{Closure}$ is the time between the initial post and the time a change request is closed (Bugzilla) or an answer has been approved in StackOverflow. We used the Mann-Whitney test to statistically compare the group means of $Time_{closure}$ for the posts including an image, versus the ones without images. Looking into the $Time_{Closure}$ showed significant difference in the group means both for StackOverflow questions (p-value = 0.01, effect size = 0.24) and Mozilla's change requests (p-value = 0.03, effect size = 0.33).

---
[3] https://www.mturk.com
[4] https://salt.bountysource.com/

**Table 1: Results of surveying the informativeness of 22,561 images using crowdsourcing with 168 developers.**

| Q1: Does the image provide additional information comparing to text? | |
|---|---|
| Yes | 87.8% |
| No | 12.2% |
| **Q2: How likely can you understand the text without the image?** | |
| Very Likely | 2.7% |
| Likely | 10.4% |
| Unlikely | 46.9% |
| Very Unlikely | 40.0% |

| Q3: What does the image communicate? | | | |
|---|---|---|---|
| Category | Overall | SO | Bugzilla |
| Code | 62.8% | 69.4% | 40.5% |
| Run time error | 42.5% | 46.2% | 30.3% |
| Menus and preferences | 17.9% | 16.9% | 21.3% |
| Dialog box | 6.7% | 1.7% | 23.3% |
| Steps and processes | 28.8% | 35.1% | 7.9% |
| Program input | 2.8% | 3.7% | 0.06% |
| Desired output | 9.2% | 10.6% | 4.6% |
| Program output | 19.1% | 20.8% | 13.7% |
| CPU/GPU performance | 2.8% | 3.18% | 1.4% |
| Algorithm/concept | 12.5% | 16.2% | 0.1% |

| Q4: Why do developers share images along with text? | |
|---|---|
| To be more precise | 70.8% |
| To be faster | 82.7% |
| To aid comprehension | 89.9% |
| Other reasons | 10.1% |

## 5 THE POSSIBLE FUTURE

**As developers (increasingly) communicate by images, we should be able to automatically retrieve, infer, and decide based on non-textual information.** Gradually, images will become a dominating means of sharing information in developers' social coding environments. A short text followed by an image will be the new format of a bug report, change request, a question, or a response. Figure 5-(i) shows a question posted by *The Passenger* on StackOverflow. Applying object recognition on the posted image (See Figure 5-(ii)) and comparing the information with the text posted by the developer, shows that there is no information within her posted text that could not be learned from the image. Indeed, the image even provided additional information in comparison to the posted text such as the Windows version, the browser used, the battery status, and the firewall status. Prospectively, *ThePassenger* will post the image plus the last sentence of her text "Did I do something wrong?". The future StackOverflow or Bugzilla, equipped with image processing capabilities, can detect and describe the problem in a structured way using image summarization and captioning methods [12]. This way, not only *ThePassenger* can post the question faster, but also, other developers can answer the question by having more details about the context extracted from the image.

As of today, tools for automated analysis of the posts in Stack-Overflow, GitHub, and Bug tracking systems only consider natural language or programming language text. Public datasets such as SOTorent only contain textual data and discard the images[5]. In consideration of the increasing trend of sharing images and the impact of social networks on developers [11], **within the near future, images will be the dominating format of communication between developers.** Initial results show that this will be in favor of making better decisions by providing more comprehensive information about the context [5]. Extending our tools to understand the content of images ultimately assist developers for more efficient communication is a predicted direction for the future.

Imagine the future that developers' productivity tools do not only provide text but also complement it with images. **Visual artifacts will be generated automatically for software teams and products.** Generating new images by synthesizing and merging parts of the former images in order to add details into artifacts is foreseeable. As a sample scenario, let's look into a bug report submitted for Mozilla's Core project within which developers discuss running Mozilla's profiler (Figure 5-(i)). Our machine learner retrieves the stored image of Mozilla's profiler (Figure 5-(ii)), which have been gathered and learned by mining Issue 131 of Firefox devtools Gecko Addon on GitHub (training set). The machine searches for the keywords on the bug report's text and matches the image with "interval" and "1ms". It then adds annotators (red rectangles) around these two texts and generated a new image that is annotated by the parts needed to be changed (Figure 5-(ii)). Comparing that with the annotated screenshot submitted by the developer, the machine-made screenshot is annotating the right parts of the image. This will help the developer in the future to submit only one bug report (originally, he submitted two bug reports).

Generating software artifacts will leverage the advances already achieved in other areas (e.g., in generating unauthentic images of humans, nature, animals, cars). We believe that within a complete learning loop, the automated tools will recommend synthesized images to developers in order to add details to their communication. Hence, the automatically generated software artifacts such as release notes, commit messages, pull request descriptions, automated responses (either chatbots, or Q/A assistance), etc. [1, 8] will include images that add detailed description of software, processes, or a task to facilitate developers' understanding by adding details.

---
[5] SoTorrent was the mining challenge dataset of MSR'18 https://github.com/sotorrent

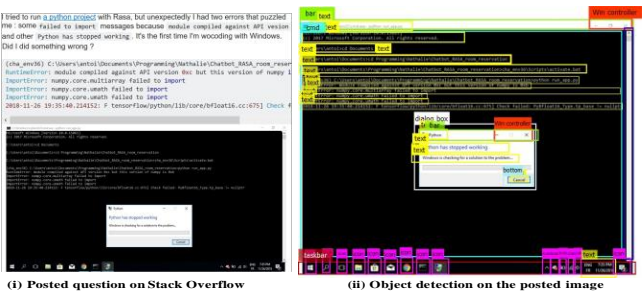

**Figure 4: Object localization and recognition**

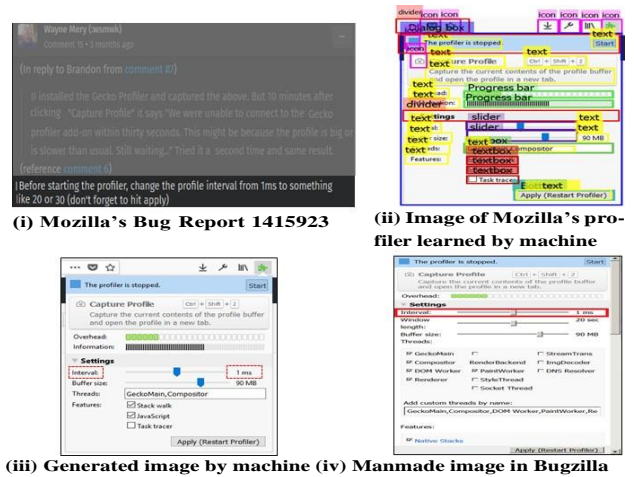

**(i) Mozilla's Bug Report 1415923**    **(ii) Image of Mozilla's profiler learned by machine**

**(iii) Generated image by machine    (iv) Manmade image in Bugzilla**

**Figure 5: An example of machine constructed output**

## 6 HOW POSSIBLE IS THIS FUTURE?

Image analysis is the process of extracting meaningful information from images by using automated techniques. Image analysis evolved into multiple sub-fields including but not limited to image recognition, image segmentation, motion detection, image captioning, and color recognition. The advances in convolutional neural networks made object detection, image captioning and summarization, generation of unauthentic images and holograms (moving images), sentiment and color analysis of images all possible and with reliable accuracy [3]. The field is growing fast with multiple applications in self driving cars, health and science, fashion, and homeland security and defence systems. While this research have been done for better understanding of the developers and their environment, we foresee that the trend of sharing images by general public would soon change the methods for extracting and analysis of software requirements [6, 7, 10].

The current body of knowledge for training image processing techniques is largely based on the annotated images of humans, nature, streets, or handwritten text [3, 12]. There is not enough data to train tools for detecting objects involved in software screenshots and artifacts. To achieve the sketched future, we need benchmark datasets to form a knowledge ontology with a variety of object classes and in a variety of text, size, color, and shape. This partonomy of images will be used to perform large scale recognition (e.g., a menu has different parts).

Software naturalness [2] conjured that most software is natural as it is developed by humans; hence is being repetitive and predictable. We argue that the job of software engineering researchers interested in pursuing our sketched future is rather more simplified in comparison to those who are focused on the processing of streets, nature images, humans or handwriting. The components and states of software are finite (already defined for the computer) and mainly known (for example, front-end developers put these known components meaningfully together). The localization of the objects (e.g., to detect a dialog box from an image, see Figure 5) is also less challenging. Software interfaces are rather structured with grid lines. Hence, the majority of objects can be defined by known geometric shapes (compared with the unknown shape of a river or a street). However, the actual performance and usability of the techniques in software engineering are subject to further analysis.

## ACKNOWLEDGMENTS

I would like to thank Guentehr Ruhe for the feedback on the early draft of this article. This research was partially supported by the Natural Sciences and Engineering Research Council of Canada, NSERC Discovery Grant RGPIN-2019-05697 and IVADO Institute for data valorization.